\documentclass[3p]{elsarticle}

\usepackage[utf8]{inputenc} 

\journal{Journal of the Mechanics and Physics of Solids}

\biboptions{authoryear}

\usepackage{color}

\usepackage{amsmath}
\usepackage{amsfonts}
\usepackage{bm} 
\usepackage{booktabs} 
\usepackage{array} 
\usepackage{paralist} 
\usepackage{verbatim} 
\usepackage{subfig} 
\usepackage[color=green!40]{todonotes}
\usepackage[colorlinks, citecolor=cyan, linkcolor=cyan]{hyperref}
\usepackage[utf8]{inputenc}

\DeclareMathOperator{\divg}{\mbox{div}} 

\newcommand{\mb}[1]{\mathbf{#1}}

\newcommand{\mc}[1]{\mathcal{#1}} 
\newcommand{\mbb}[1]{\mathbb{#1}} 

\makeatletter
\def\ps@pprintTitle{%
 \let\@oddhead\@empty
 \let\@evenhead\@empty
 \def\@oddfoot{}%
 \let\@evenfoot\@oddfoot}
\makeatother

\begin{document}
	
	\begin{frontmatter}
		
		\title{Poroelastic toughening in polymer gels: A theoretical and numerical study}
		
		\author[aff1,fn1]{Giovanni Noselli\corref{cor1}}
		\ead{giovanni.noselli@sissa.it}
		\author[aff1,fn1]{Alessandro Lucantonio}
		\author[aff2,aff3]{Robert M.~McMeeking}
		\author[aff1]{Antonio DeSimone}
		\address[aff1]{SISSA--International School for Advanced Studies, via Bonomea 265, 34136 Trieste, Italy.}
		\address[aff2]{Mechanical Engineering and Materials Departments, University of California, Santa Barbara, CA 93106, USA.}
		\address[aff3]{School of Engineering, University of Aberdeen--King's College, Aberdeen, AB24 3UE, UK.}
		\cortext[cor1]{Corresponding author.}
		\fntext[fn1]{Authors contributed equally.}
		
		\begin{abstract}
			We explore the Mode I fracture toughness of a polymer gel containing a semi-infinite, growing crack. First, an expression is derived for the energy release rate within the linearized, small-strain setting. This expression reveals a crack tip velocity-independent toughening that stems from the poroelastic nature of polymer gels. Then, we establish a poroelastic cohesive zone model that allows us to describe the micromechanics of fracture in gels by identifying the role of solvent pressure in promoting poroelastic toughening. We evaluate the enhancement in the effective fracture toughness through asymptotic analysis. We confirm our theoretical findings by means of numerical simulations concerning the case of a steadily propagating crack. In broad terms, our results explain the role of poroelasticity and of the processes occurring in the fracturing region in promoting toughening of polymer gels.
		\end{abstract}
		
		
	\end{frontmatter}


\section{Introduction}

Soft biological tissues, such as cartilage, epithelium and muscles, can sustain relatively high levels of strain, while maintaining their integrity, despite the low intrinsic toughness of the polymeric base materials \citep{Simha2004}. Since failure to withstand such loads by fracture may lead to loss of functionality and to the development of clinically adverse conditions, there is a strong biomedical interest in studying the fracture toughness of such tissues. Hydrogels are commonly considered as proxies for soft biological tissues and are thus the subject of intense theoretical and experimental investigations. Moreover, hydrogel-based technological applications in soft electronics \citep{suo_mechanics_2012}, shape morphing \citep{kempaiah_nature_2014,geryak_reconfigurable_2014}, microfluidics \citep{harmon_microfluidic_2003,kim_hydrogel-based_2007}, and in tissue engineering \citep{langer_biomaterials_2006,khademhosseini_microengineered_2007} further motivate the interest in fundamental studies on the flaw-tolerance of hydrogels and in seeking toughening mechanisms to improve their mechanical performance. 

In synthetic chemical gels, fracture occurs by scission of covalent cross-linking bonds and, as is typical of brittle materials, it usually proceeds dynamically \citep{Bonn1998,Livne2005,Livne2010}. However, the unstable character of brittle fracture may be radically altered when a brittle, impermeable hydrogel is hydraulically coupled with a tougher, poroelastic solid \citep{Casares2015,PhysRevLett.115.188105}. In such a situation, the hydraulic coupling allows for quasi-static crack propagation in the brittle solid and increases the macroscopic toughness of the composite by promoting multiple cracking \citep{noselli_analysis_2013}. More generally, toughening of hydrogels typically relies on energy dissipation in process zones around crack tips \citep{Zhao2014}. Much research currently focuses on the development of experimental protocols for the synthesis of high-toughness hydrogels, where dissipation mechanisms, such as fracture of a secondary polymer network \citep{Gong2003}, or \lq unzipping' of reversible cross-links \citep{sun_highly_2012}, are introduced at the material scale. Likewise, a theoretical effort is needed in modeling diverse fracture modes and sources of toughening in soft materials, starting with the basic, classical fracture mechanics problems.

In this spirit, here we study the propagation of a semi-infinite, parent crack in an infinite polymer gel that is loaded in Mode I conditions and immersed in a solvent. Previous theoretical works on related problems have considered stationary \citep{wang_delayed_2012,hui_stress_2013,Bouklas2015} or steadily-growing \citep{atkinson_plane_1991, radi_steady_2002} cracks, without discussing either fracture propagation criteria, or the effect of poroelastic processes in the near-tip region on the toughness of the system. Moreover, previous experimental data \citep{Tanaka2000,Lefranc2014,zhang_predicting_2015} on fracture of gels have been interpreted by invoking several dissipation sources, such as viscoelasticity, which have been quantified through phenomenological models.

In this study, we present a detailed energy analysis that reveals the existence of a velocity-independent toughening, which is innate in the poroelastic nature of polymer gels. We provide an analytical estimate of such toughening, based on asymptotic solutions of the governing equations for the crack propagation problem. As a further major result, we establish a poroelastic cohesive zone model that provides a framework to describe the micromechanics of fracture processes in polymer gels. Finally, we perform a numerical study of steady-state crack propagation, for which we explore the influence of the cohesive zone material parameters on the macroscopic toughness of the system.

\section{Governing equations for crack propagation in a polymer gel; asymptotic solutions}
\label{sec:governing}
In this section, we introduce the governing equations for Mode I crack propagation in a polymer gel, and study the asymptotic behavior of the solution near and far away from the crack tip. Specifically, we consider a gel consisting of an infinite polymer network immersed in a liquid solvent at chemical potential $\mu_{\textrm{o}} = \Omega \, p_\textrm{o}$, where $\Omega$ and $p_\textrm{o}$ are the solvent molar volume and its pressure, respectively. The gel is initially swollen with an amount of solvent that is related to the homogeneous, isotropic swelling stretch $\lambda_\textrm{o}$ from the dry state. Mechanical and chemical equilibrium conditions for the gel in its initial state imply that $\lambda_\textrm{o}$ is uniform and determined by the following equation \citep{lucantonio_transient_2013,hong_theory_2008} 
\begin{align}
\label{eq:muolamo}
\frac{RT}{\Omega}\left(\log\frac{\lambda^3_\textrm{o}-1}{\lambda^3_\textrm{o}} + \frac{1}{\lambda^3_\textrm{o}} + \frac{\chi}{\lambda_\textrm{o}^6}\right) + \frac{G_d}{\lambda_\textrm{o}} = 0 \,,
\end{align}
where $R$ is the universal gas constant, $T$ is the absolute temperature, $G_d$ is the dry shear modulus of the polymer network and $\chi$ is the dimensionless solvent-polymer mixing parameter. %
We take a plane domain $\mc{B}\subset\mbb{R}^2$ encompassing the swollen gel to be the reference configuration for all deformation of the gel network. A point in $\mc{B}$ is identified by the Cartesian coordinates $(x_1,x_2)$. Latin indices denote components of vectors and tensors with respect to the basis $\{\mb{e}_i\}, i=1,2$.

The problem to be tackled is that of a semi-infinite crack that propagates in the gel subject to applied loads capable of sustaining such crack growth. The applied loads are such that the crack tends to open with tension on the plane ahead of the crack tip and zero shear stress. The crack is thus in Mode I in the terminology of fracture mechanics. Deformations from the reference state are confined to plane-strain in conjunction with the growth of the crack.

As presented in a previous report \citep{lucantonio_reduced_2012}, the plane-strain/plane-diffusion balance equations for the linearized poroelastic model of a polymer gel read
\begin{align}
\label{eq:linbal}
&\divg{\mb{T}} = \mb{0}\,, && \dot{\varepsilon} = -\Omega\divg{\mb{h}}\,,
\end{align}
where $\mb{T}$ and $\mb{h}$ are the in-plane components of the increments of stress and solvent flux, respectively, from the reference configuration, and $\varepsilon = \mb{I}\cdot\mb{E}$ is the trace of the small-strain tensor $\mb{E}$. Recall that deformations are measured from the reference state, and thus $\varepsilon$ is the incremental volume strain relative to that state.  It is thus proportional to the solvent concentration $c$, \textit{i.e.}~$\varepsilon = \Omega c$, where $c$ defines the change of concentration relative to its value in the reference configuration, and is thus measured in moles per unit reference volume. We note that Eq.~\eqref{eq:linbal}$_1$ provides the balance of forces in the quasi-static limit and in the absence of body forces. As is usual in crack propagation problems, it is convenient to recast the governing equations in a moving frame $(\hat{x}_1,\hat{x}_2)$ centered at the crack tip, with ($r$,$\theta$) the associated polar coordinates. Spatial derivatives in the fixed and the moving frames are trivially equal and thus we use the same notation for such differential operators in both frames. In contrast, the following relation holds between the time rates of $\varepsilon$ in the moving and fixed frames
\begin{align}
\label{eq:timeder}
\varepsilon' = \dot{\varepsilon}+v\frac{\partial \varepsilon}{\partial \hat{x}_1}\,,
\end{align}
where a prime denotes the time derivative holding $(\hat{x}_1,\hat{x}_2)$ fixed and $v$ is the crack tip velocity along $\mb{e}_1$. Relationships similar to eq.~\eqref{eq:timeder} hold for additional fields such as displacement.
 
The governing equations include constitutive laws. First, for the relationship between the solvent flux and the driving force for solvent migration, \textit{i.e.} the gradient of chemical potential $\mu$, we choose Darcy's law: $\mb{h}=-M\nabla \mu$, where $M$ is the solvent mobility and $\mu$ is the incremental chemical potential added to $\mu_{\textrm{o}}$. Then, for the stress-strain relationships we prescribe the following plane strain poroelasticity equations, valid for a homogeneous, isotropic polymer gel,
\begin{align}
\label{eq:const}
\mb{T} = 2G\,\mbox{dev}\,\mb{E}+\left(\kappa\varepsilon - \frac{\mu}{\Omega}\right)\mb{I}\,, && \mbox{dev}\,\mb{E} = \mb{E}-\frac{\varepsilon}{3}\mb{I}\,,
\end{align}
with the following definitions of the poroelastic moduli \citep{lucantonio_reduced_2012}
\begin{align}
\label{eq:poromod}
&G = \frac{G_d}{\lambda_\textrm{o}}\,, &&\kappa = -\frac{G}{3}+\frac{R T}{\Omega}\frac{2\chi+(1-2\chi)\lambda_\textrm{o}^3}{\lambda_\textrm{o}^6(\lambda_\textrm{o}^3-1)}\,.
\end{align}
We observe that $\mu/\Omega$ is the pressure in the solvent and thus the terms in Eq.~\eqref{eq:const} other than that is the stress sustained by the polymer network. We also notice that, apart from the definitions Eq.~\eqref{eq:poromod}, the governing equations of the linearized model for swelling polymer gels are the same as those of the Biot theory of poroelasticity \citep{biot_general_1941,RiceCleary1976} in the presence of incompressible solid and fluid constituents.

To complete the set of governing equations, we specify the boundary conditions by assuming that the crack faces are traction-free ($\mb{Tn}=\mb{0}$, with $\mb{n}$ the outwards unit normal to the faces) and in chemical equilibrium with the initial/reference state of the solvent ($\mu=0$). At infinity, the Mode I stress field is applied, such that
\begin{align}
\label{eq:modeIsol}
T_{ij}(r,\theta)  =  \frac{K_{\textrm{I}}}{\sqrt{2\pi r}}\cos\frac{\theta}{2}\, f_{ij}(\theta)\,,
\end{align}
where the stress intensity factor has the value $K_v^\infty$ and
\begin{align}
&f_{11} = 1-\sin\frac{\theta}{2}\sin\frac{3\theta}{2}\,, && \quad f_{12} = \sin\frac{\theta}{2}\cos\frac{3\theta}{2}\,, && f_{22} = 1+\sin\frac{\theta}{2}\sin\frac{3\theta}{2}\,.
\end{align}
The applied stress intensity factor $K_v^\infty$ depends, in general, on crack tip velocity and is determined by the crack propagation criterion, as discussed later in Section~\ref{sec:energy}.
Moreover, at infinity the material responds as an incompressible elastic solid, under the assumption that  fluid flow is confined to a small zone surrounding the crack tip. We call this condition the {\it small-scale process zone} hypothesis. Note that for a finite-size gel specimen, this hypothesis holds as long as the characteristic size of the process zone is much smaller than the size of the specimen. Since the local volume change, upon taking the trace of Eq.~\eqref{eq:const}, is
\begin{align}
\label{eq:traceconst}
\varepsilon = \frac{1}{\kappa + G/3}\left(\frac{1}{2}\mb{I}\cdot\mb{T} + \frac{\mu}{\Omega}\right)\,,
\end{align}
the incompressibility assumption $\varepsilon = 0$ for $r\rightarrow \infty$ implies that the far-field pressure is given by
\begin{align}
\label{eq:pressinc}
\frac{\mu}{\Omega}(r,\theta) = -\frac{1}{2}\mb{I}\cdot\mb{T} = -\frac{K_v^\infty}{\sqrt{2\pi r}}\cos\frac{\theta}{2}\,.
\end{align}

The problem may be formulated using stresses as the primary unknowns; to this aim, we introduce the Airy stress function  $\phi$ such that
\begin{align}
T_{11} = \frac{\partial^2 \phi}{\partial \hat x_2^2}\,, && T_{22} = \frac{\partial^2 \phi}{\partial \hat x_1^2}\,, && T_{12} = -\frac{\partial^2 \phi}{\partial \hat x_1 \partial \hat x_2}\,,
\end{align}
satisfying the balance of forces Eq.~\eqref{eq:linbal}$_1$ identically. In order to be able to find the displacement field by integration of the strains we impose the plane strain compatibility condition 
\begin{align}
\frac{\partial^2E_{11}}{\partial \hat x_2^2} + \frac{\partial^2 E_{22}}{\partial \hat x_1^2} = 2\frac{\partial^2 E_{12}}{\partial \hat x_1 \partial \hat x_2}\,,
\end{align}
written in terms of stresses by using Eq.~\eqref{eq:const} as
\begin{align}
\label{eq:biharmonic}
\Delta\Delta \phi + \frac{6 G}{4G+3\kappa}\frac{\Delta \mu}{\Omega} = 0\,,
\end{align}
with $\Delta()=\partial^2()/\partial \hat x_1^2 + \partial^2()/\partial \hat x_2^2$,  $\Delta\Delta()=\partial^4()/\partial \hat x_1^4 + 2\partial^4()/\partial \hat x_1^2 \partial \hat x_2^2 + \partial^4()/\partial \hat x_2^4$.

With this, the governing equations of the problem in terms of $\phi$ and $\mu$, written in the reference frame convected with the crack tip, are
\begin{align}
\label{eq:balaeq}
&\Delta\Delta \phi + \frac{6 G}{4G+3\kappa}\frac{\Delta \mu}{\Omega} = 0\,, && \varepsilon'-v\frac{\partial \varepsilon}{\partial \hat{x}_1} = M\Omega \Delta \mu\,,
\end{align}
where $\varepsilon$ is given by Eq.~\eqref{eq:traceconst}. By expressing $\Delta \mu$ as a function of $\Delta \varepsilon$ through Eqs.~\eqref{eq:balaeq}$_1$ and \eqref{eq:traceconst}, we may recast Eq.~\eqref{eq:balaeq}$_2$ in the form of an advection-diffusion equation
\begin{align}
\label{eq:diffadv}
\varepsilon'-v\frac{\partial \varepsilon}{\partial \hat{x}_1} = D_c \Delta \varepsilon\,,
\end{align} 
with $D_c = M\Omega^2(4G/3+\kappa)$ the cooperative diffusion coefficient \citep{hui_stress_2013}. This equation involves two length scales: $\sqrt{D_c t}$ and $l=D_c/v$. Both lengths characterize, in different regimes and under the small scale process zone hypothesis, the extent of the process zone, where significant volume changes occur. Specifically, the length scale $\sqrt{D_c t}$ is relevant at short times for a stationary crack, whereas $l$ becomes relevant during crack propagation. 

\subsection{Asymptotic solutions}\label{sec:ass}

The formulation of the governing equations as in \eqref{eq:balaeq} and \eqref{eq:diffadv} enables insights from asymptotic analysis. At the crack faces, the solvent pressure must be zero to satisfy the boundary condition on chemical potential. For $r \rightarrow 0$,  the stress may freely develop a singularity, whereas we assume a non-singular pressure profile. The instantaneous relaxation of pore pressure at the crack tip motivates our assumption \citep{hui_stress_2013}, which is confirmed by numerical results. Thus, the leading order form of the balance equations \eqref{eq:balaeq} is
\begin{align}
\label{eq:balalead}
&\Delta\Delta \phi = 0\,, && \varepsilon'-v\frac{\partial \varepsilon}{\partial \hat{x}_1} = M\Omega \Delta \mu\,,
\end{align}
where $\varepsilon$ is given by the leading order form of Eq.~\eqref{eq:traceconst}
\begin{align}
\label{eq:traceconstlead}
\varepsilon = \frac{1}{\kappa + G/3}\frac{1}{2}\Delta \phi\,.
\end{align}
The problem becomes one-way coupled, since we can solve Eq.~\eqref{eq:balalead}$_1$ for $\phi$ (subject to the traction boundary conditions on the crack faces), and then recover $\mu$ from Eq.~\eqref{eq:balalead}$_2$, once $\varepsilon$ is known from the stresses through Eq.~\eqref{eq:traceconstlead}. First, the asymptotic solution of the compatibility equation \eqref{eq:balalead}$_1$ for $r \rightarrow 0$ is the Mode I stress field Eq.~\eqref{eq:modeIsol}, with $K_{\textrm{I}}=K_{\textrm{tip}}$, the stress intensity factor at the crack tip, to be determined by the crack propagation criterion. Then, from Eq.~\eqref{eq:traceconstlead}
\begin{align}
\label{eq:traceasy}
\varepsilon = \frac{1}{\kappa + G/3} \frac{K_{\textrm{tip}}}{\sqrt{2\pi r}}\cos \frac{\theta}{2}\,,
\end{align}
so that the rate of change of the dilatational strain is
\begin{align}
\label{eq:dilatationrate}
\varepsilon'-v\frac{\partial \varepsilon}{\partial \hat{x}_1} = \frac{1}{\kappa + G/3}\left(\frac{\partial K_{\textrm{tip}}}{\partial t} + \frac{v K_{\textrm{tip}}}{2 r}\right)\frac{1}{\sqrt{2\pi r}}\cos \frac{\theta}{2}\,.
\end{align}
Inserting this expression into Eq.~\eqref{eq:balalead}$_2$ and solving the corresponding problem \citep{atkinson_plane_1991} for $\mu$, while retaining only the most singular contribution, we have at leading order
\begin{align}
\label{eq:asypresstip}
\frac{\mu}{\Omega} = \left(-\frac{v}{4M\Omega^2 (\kappa + G/3)}\frac{K_{\textrm{tip}}}{\sqrt{2\pi}}\cos{\frac{3\theta}{2}}+K_2 \cos{\frac{\theta}{2}}\right)\sqrt{r}\,,
\end{align}
where $K_2$ depends on the applied stress intensity factor $K^\infty_v$ at infinity. We note that this result still holds for a stationary crack, for which the first term in the parenthesis of Eq.~\eqref{eq:dilatationrate} must be considered. In a previous work \citep{hui_stress_2013}, the stationary crack problem  has been studied in detail.

\begin{figure}[!th]
	\centering
	\includegraphics[scale=1]{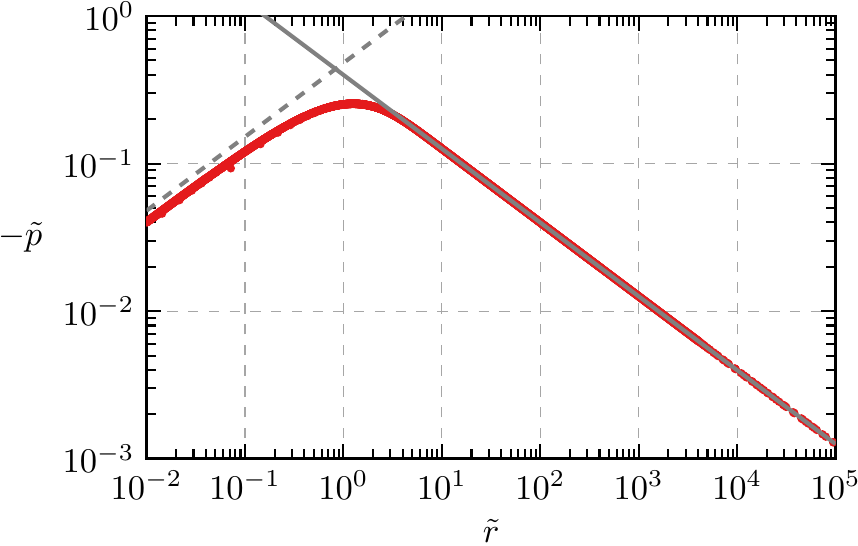}
	\caption{Dimensionless solvent negative pressure $-\tilde{p} = -\mu\sqrt{l}/(\Omega K_v^\infty)$ along the line $\theta = 0$ as a function of the dimensionless distance $\tilde{r} = r/l$ from the crack tip. Numerical results (solid red curve) obtained with the energy release rate approach (see Sections~\ref{sec:energy} and \ref{sec:results} for the derivation of the energy release rate and the details about the numerical solution). The dashed grey line is the asymptotic behavior $\sim\sqrt{r}$ predicted theoretically by Eq.~\eqref{eq:asypresstip} near the crack tip. The continuous grey line is the inverse square-root behavior given by Eq.~\eqref{eq:pressinc}, far away from the crack tip.}
	\label{fig:adimpress}
\end{figure}

For $r \gg l$, the gradient of the volume change is negligible, which implies, for the conditions at infinity, that the response of the material is undrained, \textit{i.e.} $\varepsilon=0$. Hence, from Eq.~\eqref{eq:traceconst}, 
\begin{align}
	\label{eq:muinc}
	\frac{\mu}{\Omega} = -\frac{1}{2}\Delta\phi\,,
\end{align}
so that Eq.~\eqref{eq:biharmonic} reduces to $\Delta \Delta \phi = 0$. Again, the solution of the compatibility equation is the Mode I stress field Eq.~\eqref{eq:modeIsol}, with $K_{\textrm{I}}=K_v^\infty$ to satisfy the conditions at infinity, from which, by Eq.~\eqref{eq:muinc}, $\mu$ is harmonic and given explicitly by Eq.~\eqref{eq:pressinc}.

We observe that $K_2$ in Eq.~\eqref{eq:asypresstip} is such that the term in parenthesis is negative, in order for the asymptotic solution to match the far-field solution of Eq.~\eqref{eq:pressinc}. With this, by computing the gradient of the solvent chemical potential from Eq.~\eqref{eq:asypresstip}, it may be verified that these terms will induce a circulating flow in which solvent is pumped from the crack surfaces, injected round towards the region ahead of the crack tip and dumped into the crack tip \citep{wang_delayed_2012}. This circulating flow is convected through the gel by the growth of the crack.

We confirm this asymptotic analysis by carrying out numerical simulations for the steady-state case of constant crack tip velocity $v$, as described later in Section~\ref{sec:results}. As depicted in Fig.~\ref{fig:adimpress}, the solvent pressure decreases as $\sqrt{r}$ according to Eq.~\eqref{eq:asypresstip} ahead of the crack tip, reaches a minimum at $r \approx D_c/v = l$, and then increases again, following the scaling law Eq.~\eqref{eq:pressinc}. The minimum of $\tilde{p}$ approximately marks the boundary between the drained and the undrained regions, both for a propagating and for a stationary \citep{hui_stress_2013} crack. 

\section{Energetics of crack propagation in polymer gels}
\label{sec:energy}

In the following, we derive and elaborate on the energy release rate for a polymer gel, a fundamental quantity for establishing crack propagation criteria. Although the technical aspects of the derivation are rather standard in fracture mechanics, some special aspects must be considered that stem from the poroelastic nature of gels. In particular, energy dissipation due to solvent transport gives rise to an area integral in the expression of the energy release rate, in addition to the classical contour integral, which critically contributes to the fracture toughness of the gel. In general, we find it useful to dwell on some key intermediate steps of the derivation to comment on their physical significance.

\begin{figure}[!th]
	\centering
	\includegraphics[scale=1]{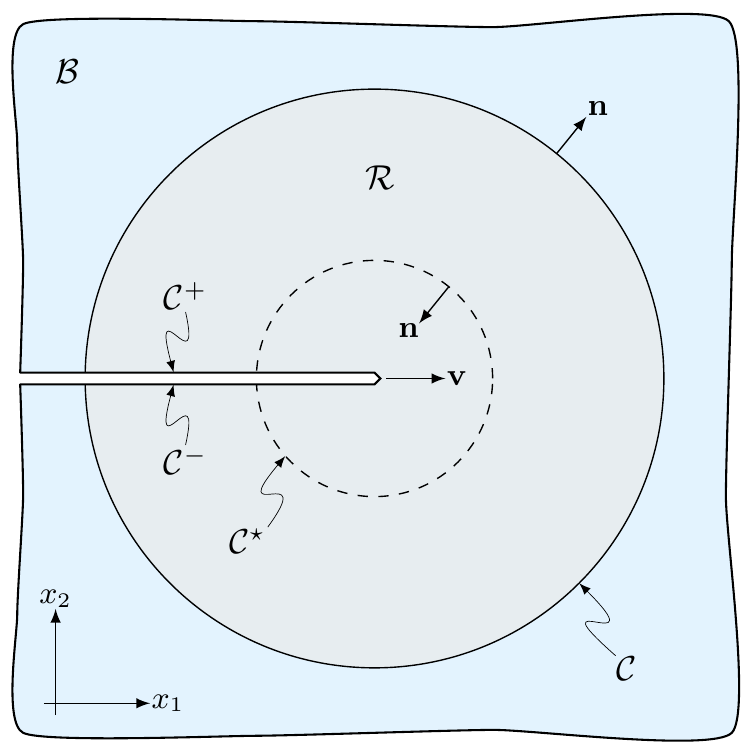}
	\caption{Sketch for the computation of the energy release rate. The material domain $\mc{R}\subset\mc{B}$ includes the crack tip and is enclosed by the contour $\mc{C}$ joining the crack faces and including the two segments $\mc{C}^+$ and $\mc{C}^-$ along them.}
	\label{fig:fig2}
\end{figure}

Let us consider a material domain $\mc{R}\subset\mc{B}$ including the crack tip and enclosed by the contour $\mc{C}$ that joins the crack faces and includes the two segments $\mc{C}^+$ and $\mc{C}^-$ along the crack faces, see Fig.~\ref{fig:fig2}. In an isothermal setting, we consider power balances per unit thickness of the body.

The external power consists of $P_{m}$, the rate at which the applied tractions do work on $\partial{\mc{R}}=\mc{C}\cup\mc{C}^+\cup\mc{C}^-$, and of $P_{ch}$, the rate at which solvent transport convects energy across $\partial{\mc{R}}$:
\begin{align}
\label{eq:pext}
&P_{m}(\partial\mc{R}) = \int_{\partial\mc{R}}{\mb{Tn}\cdot\dot{\mb{u}}\,\mbox{d}S}\,, 
&& P_{ch}(\partial\mc{R}) = -\int_{\partial\mc{R}}{\mu\,\mb{h}\cdot\mb{n}\,\mbox{d}S}\,,
\end{align}
where $\mb{n}$ is the outward unit normal to $\partial\mc{R}$ and $\dot{\mb{u}}$ is the velocity of the gel. These representations are appropriate when the polymer gel is viewed as a homogenized continuum, where no distinction is made between polymer and solvent. However, we recognize that there is a clear connection with mixture theory, as the solvent flux $\mb{h}$ is proportional to the relative velocity of the solvent with respect to the polymer and the total stress $\mb{T}$ is the sum of the network stress $\mb{T}^n$ and a hydrostatic contribution $-(\mu/\Omega)\mb{I}$ equal to the solvent pressure. As can be deduced from Eqs.~\eqref{eq:const} and \eqref{eq:poromod}, the network stress comprises an elastic component, which expresses the elasticity of the cross-linked polymers, and an osmotic component, which arises from the mixing of solvent and polymers and is present even without cross-links \citep{sekimoto_role_1994}. In particular, the network traction $\mb{T}^n\mb{n}$ is work-conjugate to the network velocity $\dot{\mb{u}}$, while the solvent pressure $\mu/\Omega$ does work on the solvent velocity $\mb{v}_s = \dot{\mb{u}}+\Omega \mb{h}$. Thus, we may alternatively represent the total external power as
\begin{align}
\int_{\mc{\partial \mc{R}}}{\left(\mb{T}^n\mb{n}\cdot\dot{\mb{u}}-\frac{\mu}{\Omega}\mb{n}\cdot\mb{v}_s\right)\mbox{d}S}\,.
\end{align}
This result makes it clear that Eq.~\eqref{eq:pext}$_2$ is the rate at which the tractions do work on the fluid velocity in excess of the network velocity.
The external power contributes to the change of the free energy $E$ stored in $\mc{R}$ and to the dissipation $D$ due to solvent transport:
\begin{align}
\label{eq:timerateen}
&\dot{E}(\mc{R}) = \frac{\mbox{d}}{\mbox{d}t}\int_{\mc{R}}{\psi\,\mbox{d}A}\,, && 
D(\mc{R}) = -\int_{\mc{R}}{\mb{h}\cdot\nabla\mu\,\mbox{d}A}\,;
\end{align}
the remainder, denoted by $\Phi$, is directed into the crack tip and is the driver for crack propagation. Here, the free energy density $\psi$ is defined as
\begin{align}
\psi(\mb{E})  = G\,\textrm{dev}\mb{E}\cdot\textrm{dev}\mb{E}+\frac{1}{2}\kappa\varepsilon^2\,,
\end{align}
so that
\begin{align}
\mb{T}^n = \frac{\partial \psi}{\partial \mb{E}}\,.
\end{align}
With the aforementioned definitions of the power contributions, we write the balance of energy for $\mc{R}$ as
\begin{align}
P_{m}(\partial\mc{R}) + P_{ch}(\partial\mc{R}) = \dot{E}(\mc{R}) + D(\mc{R}) + \Phi\,. \label{eq:powerbal}
\end{align}
Consistent with the hypothesis of quasi-static crack propagation, we have neglected the kinetic energy in the energy balance.

Manipulations of the general balance statement Eq.~\eqref{eq:powerbal} lead to the definition of the energy release rate. First, we recall the following transport theorem \citep{gurtin_configurational_1996}
\begin{align}
\dot{E}(\mc{R})=\int_{\mc{R}}{\psi'\,\mbox{d}A}-\int_{\partial{\mc{R}}}{\psi v n_1\,\mbox{d}S}\,,
\end{align}
where $n_1 = \mb{n}\cdot\mb{e}_1$.
With this, we can compute the power $\Phi$ dissipated at the crack tip from Eq.~\eqref{eq:powerbal} as
\begin{align}
	\label{eq:tipdiss}
	\Phi =  \int_{\mc{\mc{C}}}{(\psi v n_1+\mb{Tn}\cdot\dot{\mb{u}}-\mu\mb{h}\cdot\mb{n})\,\mbox{d}S} - \int_{\mc{R}}{(-\mb{h}\cdot\nabla\mu+\psi')\mbox{d}A}\,,
\end{align}
where we have taken into account the fact that the boundary conditions on the crack faces imply no exchange of either mechanical or chemical power. Clearly, Eq.~\eqref{eq:tipdiss} shows that only part of the energy flux across $\mc{C}$ is directed towards the tip, since part is dissipated in solvent transport, and part is spent in transient processes that vary the free energy stored in $\mc{R}$.

On applying the divergence theorem  \citep{gurtin_configurational_1996} to the chemical power contribution in Eq.~\eqref{eq:tipdiss} we obtain
\begin{align}
\label{eq:chempowdiv}
	-\int_{\mc{C}}{\mu\mb{h}\cdot\mb{n}\,\mbox{d}S} = -\oint_{\textrm{tip}}{\mu\mb{h}\cdot\mb{n}\,\mbox{d}S} + \int_{\mc{R}}{\left(\frac{\mu}{\Omega}\dot{\varepsilon}-\mb{h}\cdot\nabla\mu\right)\mbox{d}A}\,,
\end{align}
where the first term at the right hand side is the limit of the integral over a circular contour that is shrunk to the crack tip. This term vanishes because the asymptotic solution Eq.~\eqref{eq:asypresstip} makes it the integral of a bounded quantity on a vanishingly small contour: $\mu \sim \sqrt{r}$ and $\mb{h} \sim 1/\sqrt{r}$. Furthermore, upon noting that $\psi' = \mb{T}^n\cdot\mb{E}' = \mb{T}^n\cdot\nabla\mb{u}'$, we see that the last term on the right hand side of Eq.~\eqref{eq:tipdiss} becomes
\begin{align}
\label{eq:nonsteadypsi}
\int_{\mc{R}}{\psi'\,\mbox{d}A} = -\oint_{\textrm{tip}}{\mb{Tn}\cdot\mb{u}'\,\mbox{d}S} + \int_{\mc{C}}{\mb{Tn}\cdot\mb{u}'\,\mbox{d}S}+\int_{\mc{R}}{\frac{\mu}{\Omega}\varepsilon'\,\mbox{d}A}\,,
\end{align}
where we have used the divergence theorem and the balance of forces. Note that the tip integral vanishes for the Mode I stress field at the crack tip. Then, from Eqs.~\eqref{eq:chempowdiv}-\eqref{eq:nonsteadypsi} and the transformation rule Eq.~\eqref{eq:timeder} applied to $\dot{\mb{u}}$ and $\dot{\varepsilon}$, Eq.~\eqref{eq:tipdiss} simplifies to 
\begin{align}
	\label{eq:Phisteady}
	\mc{G} = \frac{\Phi}{v}  =  \int_{\mc{C}}{\left(\psi n_1 - \mb{T}\mb{n}\cdot\frac{\partial\mb{u}}{\partial \hat{x}_1}\right)\mbox{d}S} - \int_{\mc{R}}{\frac{\mu}{\Omega}\frac{\partial \varepsilon}{\partial \hat{x}_1}\,\mbox{d}A}\,,
\end{align}
where we have defined the \textit{energy release rate} $\mc{G}$.
This expression is consistent with that obtained in \cite{Bouklas2015} following a different derivation and in the finite deformation context. Notice that $\Phi$ is path-independent since the choice of the path $\mc{C}$ enclosing $\mc{R}$ is arbitrary. In particular, we may choose a contour that is shrunk to the crack tip, such that the area integral vanishes due to an argument similar to that used in the discussion of the tip integral in Eq.~\eqref{eq:chempowdiv}. Thus, we may rewrite Eq.~\eqref{eq:Phisteady} as
\begin{align}
\label{eq:G}
\mc{G} = \oint_{\textrm{tip}}{\left(\psi n_1-\mb{T}^n\mb{n}\cdot\frac{\partial\mb{u}}{\partial \hat{x}_1}\right)\mbox{d}S}\,.
\end{align}
This representation for the energy release rate shows that an infinitesimal contour surrounding the crack tip is needed to exclude the contribution of the solvent pressure in the computation of $\mc G$. Thus, only the work performed by the network stress and the energy stored by the network are involved in the energy balance at the crack tip. 
Upon using the asymptotic stress field of Eq.~\eqref{eq:modeIsol} for $r\rightarrow 0$  with $K_{\textrm{I}}=K_{\textrm{tip}}$ and the corresponding strain field, we compute the energy release rate explicitly from Eq.~\eqref{eq:G} as
\begin{align}
\label{eq:Gtip}
\mc{G}=\frac{K_{\textrm{tip}}^2}{\bar{E}}\,,
\end{align}
with $\bar{E}=E/(1-\nu^2)$ the drained plane strain modulus of the gel, defined as a function of the drained elastic moduli
\begin{align}
E = \frac{9\kappa G}{3\kappa + G}\,, && \nu = \frac{3\kappa-2G}{2(3\kappa + G)}\,.
\end{align}

As a simple fracture criterion, we may assume that the energy $\mc{G}$ absorbed into the crack tip is spent in the breaking of polymer chains lying across the crack plane. This behavior can be characterized by a material-dependent fracture energy $\Gamma$ per unit area, and thus the fracture criterion is $\mc{G}=\Gamma$, {\it i.e.} $K_{\textrm{tip}} = \sqrt{\Gamma \bar{E}} = K^\infty_\textrm{o}$. Here, $K^\infty_\textrm{o}$ is the critical stress intensity factor needed to commence the propagation of a stationary crack in the absence of solvent pressure. Since the area density of the polymer chains is a function of the degree of swelling of the gel, the fracture energy depends on the swelling stretch $\lambda_\textrm{o}$ \citep{Zhao2014}. 

\begin{figure}[!t]
	\centering
	\includegraphics[scale=1]{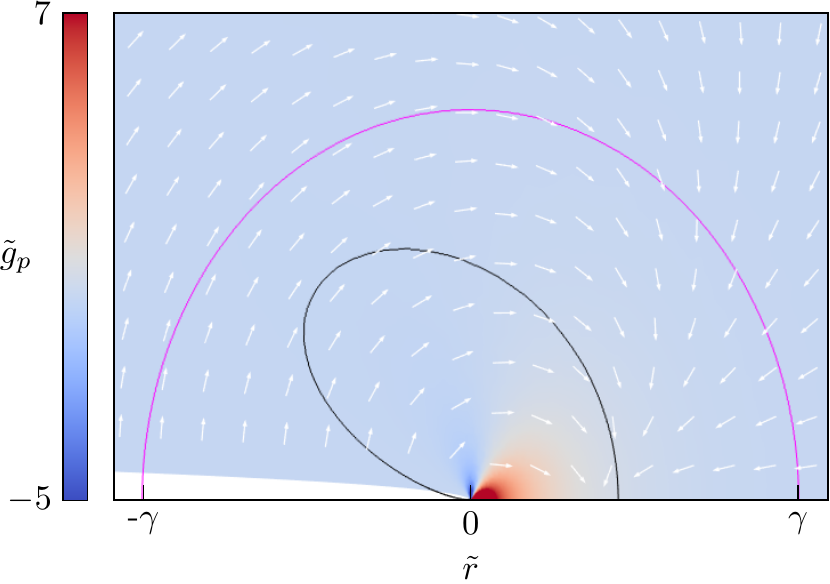}
	\caption{Close-up view of the crack tip region with arrows indicating the solvent flux field $\mb{h}$. Color code represents the dimensionless power density $\tilde{g}_p = (\mu l\partial \varepsilon/\partial \hat{x}_1)/G\Omega$, as computed from numerical simulations (see Section~\ref{sec:results}), for $\kappa/G=10$ and $\Gamma/G l\approx 60$. The magenta contour $\tilde r = \gamma$, with $\gamma \approx 3.4$, is an estimate for the boundary of the process zone, where $\gamma$ has been tuned to match the value of $K_v^\infty$ computed from Eq.~\eqref{eq:Kinfty} with the corresponding value obtained numerically. Outside the black contour, which is the level set $\varepsilon=0.1$, volume changes are less than 10\%, and the material response approaches the undrained one.}
	\label{fig:processzone}
\end{figure}

As we have seen in the previous section, during crack propagation solvent flow is confined to a small region near the crack tip. Outside this region, the incompressible Mode I stress field holds, with the solvent pressure given by Eq.~\eqref{eq:pressinc}. As a consequence, the first term in Eq.~\eqref{eq:Phisteady}, when evaluated on a contour $\mc{C}^\infty$ outside the process zone, gives the usual relation $\mc{G}_v^\infty = (K_v^\infty)^2/\bar{E}_u$, where $\mc{G}_v^\infty$ is the applied energy release rate and $\bar{E}_u=4E/3=4G$ is the undrained plane strain modulus. 
Then, from the comparison of Eq.~\eqref{eq:Phisteady} and Eq.~\eqref{eq:Gtip}, we deduce that the energy release rate computed in the far field may be expressed as
\begin{align}
\label{eq:Gexp}
\mc{G}_v^\infty = \Gamma + \int_{\mc{R}^\infty}{\frac{\mu}{\Omega} \frac{\partial \varepsilon}{\partial \hat{x}_1}}\,\mbox{d}A\,.
\end{align}
This is a central result of our work that applies to both a stationary and a propagating crack, as long as the small-scale process zone hypothesis holds true. The area integral in Eq.~\eqref{eq:Gexp}, extending over the domain $\mc{R}^\infty$ encircled by $\mc{C}^\infty$, effectively involves only the process zone of characteristic size $\gamma l$, with $\gamma$ a positive constant. We confirm this observation by plotting the dimensionless power density $\tilde{g}_p = (\mu l\partial \varepsilon/\partial \hat{x}_1)/G\Omega$. This is reported in Fig.~\ref{fig:processzone} as computed from numerical simulations, see Section~\ref{sec:results} for additional details. Thus, using the asymptotic fields Eqs.~\eqref{eq:traceasy}-\eqref{eq:asypresstip}, we may estimate the area integral in Eq.~\eqref{eq:Gexp} as
\begin{align}
\int_{0}^{2\pi}\int_{0}^{\gamma l}{\frac{\mu}{\Omega} \frac{\partial \varepsilon}{\partial \hat{x}_1}\,r\,\mbox{d}r \mbox{d}\theta} = \frac{3}{16}\gamma\frac{4G+3\kappa}{(G+3\kappa)^2}K_{\textrm{tip}}^2 \, ,
\end{align}
and solve Eq.~\eqref{eq:Gexp} for $K^\infty_v$ to obtain
\begin{align}
\label{eq:Kinfty}
K^\infty_v = K^\infty_\textrm{o}  \sqrt{ \frac{4}{3}(1-\nu^2)\left[1 + \frac{\gamma}{4}(1-2\nu)\right] }\, .
\end{align}
Interestingly, Eq.~\eqref{eq:Kinfty} shows that the stress intensity factor $K^\infty_v$ needed to sustain crack growth, that is, the effective fracture toughness of the system, is independent of crack tip velocity and is solely a function of the Poisson's ratio $\nu$. We have confirmed this result by means of numerical simulations that also allowed the estimation of the constant $\gamma$.
Physically, this toughening effect may be explained as follows. The stress concentration ahead of the crack tip causes the material to expand. Since expansion requires an increase in solvent volume fraction, a reduction of solvent pressure develops in the process zone to draw solvent towards the crack tip. This solvent flow implies viscous dissipation. In fact, energy must be supplied to the system to compensate for the work performed on solvent pressure, which resists volume expansion within the process zone, as quantified by the area integral in Eq.~\eqref{eq:Gexp}. 

Notice that the degree of toughening $K^\infty_v/K^\infty_{\textrm{o}}$ is a function of the initial swelling stretch $\lambda_{\textrm{o}}$ through the poroelastic moduli, see Eq.~\eqref{eq:poromod}. As may be expected, the toughening contribution in Eq.~\eqref{eq:Kinfty} vanishes for an incompressible material, \textit{i.e.}~a dry polymer gel. This fact is consistent with the physical interpretation of the toughening as a dilatational effect, since volume changes in the process zone will be hampered by incompressibility.   

In passing, we observe that a related result was previously obtained by \cite{hui_stress_2013} for the case of a Mode I stationary crack in a polymer gel. Consistent with our findings, the enhancement in fracture toughness as predicted by these authors  vanishes at $\nu = 1/2$.

\subsection{Domain integral method}

To accurately compute the energy release rate $\mc{G}$ in numerical simulations, we adopt the domain integral method \citep{Li1985}. We consider a closed path comprising the contours $\mc{C}^\star$ and $\mc{C}$ encircling the crack tip, and the two segments along the crack faces joining them, see Fig.~\ref{fig:fig2}. Moreover, we define $q$ to be a smooth function of position such that $q=0$ on $\mc{C}$ and $q=1$ on $\mc{C}^\star$. With these definitions, we rewrite Eq.~\eqref{eq:G} as
\begin{align}
	\mc{G} = -\lim_{\mc{C}^\star \rightarrow 0} \int_{\mc{C}\cup\mc{C}^\star}{q\left(\psi n_1-\mb{Tn}\cdot\frac{\partial\mb{u}}{\partial \hat{x}_1}\right)\mbox{d}S} = -\mb{e}_1\cdot\int_{\mc{R}}{\divg (q\mbb{E})\,\mbox{d}A}\,,
\end{align}
where $\mbb{E} = \psi\mb{I}-(\nabla^{\textrm{T}}\!\mb{u})\mb{T}$ is the Eshelby tensor \citep{gurtin_energy_1979} and the divergence theorem has been applied in the last equality. Using the constitutive equations and the balance of forces \eqref{eq:linbal}$_1$ we have $\divg\mbb{E} = (\mu/\Omega)\nabla\varepsilon$, so that
\begin{align}
	\label{eq:Jdomint}
	\mc{G} = -\int_{\mc{R}}{\left(\mbb{E}\nabla q \cdot \mb{e}_1 + q\frac{\mu}{\Omega} \frac{\partial \varepsilon}{\partial \hat{x}_1}\right)\mbox{d}A}\,.
\end{align}

\section{Poroelastic cohesive zone model}
\label{sec:cohzone}

In this section, we establish a poroelastic cohesive zone model that takes both deformation of the polymer network and solvent transport into account. This micromechanical approach allows us to distinguish among several failure mechanisms that otherwise cannot be accounted for by a fracture criterion based solely on the crack tip toughness.  

In the cohesive zone adjacent to the crack tip the polymer network fails by progressive rupture of its chains. This process is accompanied by two phenomena, one of which is dissipation due to friction and sliding to untangle the ruptured polymer chains; this phenomenon means that the work to rupture the network of cross-linked polymer is much higher than the energy required to simply break its bonds. The second phenomenon is the flow of solvent to fill the space created by the rupturing of the cohesive zone, whose opening is zero where rupture begins and rises to a maximum where the polymer network finally fails. The flow of solvent involves viscous drag and therefore introduces further dissipation that will depend on the rate at which the crack propagates.

\begin{figure}[!h]
	\centering
	\includegraphics[scale=1]{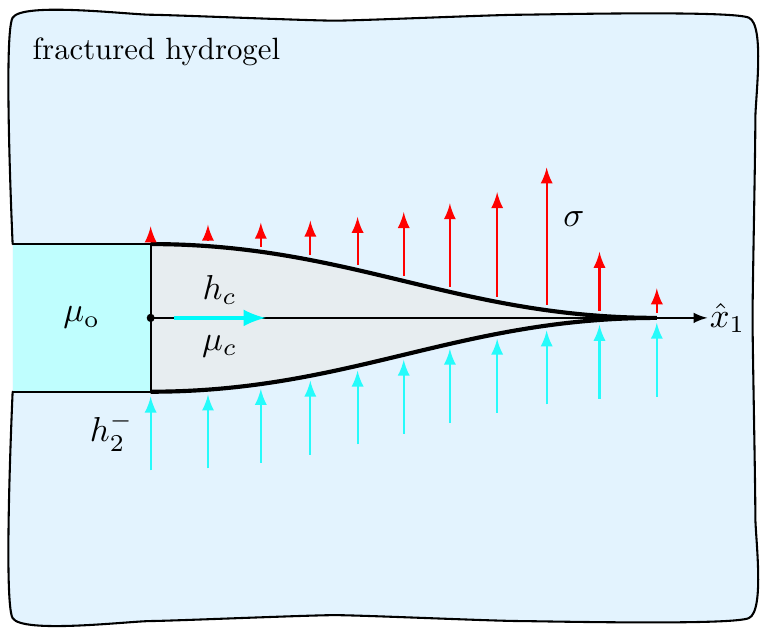}
	\caption{A sketch of the poroelastic cohesive zone adjacent to the crack tip. Note that the cohesive model takes both the deformation of the polymer network and the transport of solvent into account. Here, $\sigma$ is the cohesive traction, $h_c$ is the tangential flux of solvent within the cohesive zone, whereas $h_2^-$ and $h_2^+$ are the fluxes exchanged with the surrounding gel.}
	\label{fig:cohzone}
\end{figure}

In a precise and complete treatment of the problem, it would be possible to extend the general features of flow in the gel as treated in Section~\ref{sec:governing} to encompass the phenomenon of flow into the rupturing cohesive zone.  Instead of doing so, we prefer to treat flow in the cohesive zone as a separate feature that is in equilibrium and compatible with the surrounding gel.  By treating flow in this manner, we will be able to obtain additional insights into its contribution to the toughness of the gel.  Specifically, the work done on the cohesive zone, which is required to rupture the polymer network and to force the solvent into the space created by the failure process, is additional to the elastic deformation of the polymer network and the associated dissipation due to solvent flow. 

We now state the governing equations for the cohesive zone. We notice that, due to the planar nature of the present setting, the cohesive zone coincides with the interval $0\leq \hat{x}_1 \leq R_v$, where $R_v$ is its length, generally depending on $v$. The balance of forces prescribes that the stresses $T^{+}_{22}$ and $T_{22}^{-}$ exerted by the cohesive zone on the gel along the planes $\hat{x}_2 = 0^+$ and $\hat{x}_2 = 0^-$, respectively, must equal the traction $\sigma$ within the cohesive zone. We have assumed that the tractions exchanged between the gel and the cohesive zone only have the component along $\mb{e}_2$, consistent with a Mode I crack.

To enable analysis of the solvent pressure in the cohesive zone, we develop a transport model for the solvent within it. We assume that the cohesive zone permits solvent flux $h_c$ (measured in $\mbox{mol}/\mbox{m}\cdot\mbox{s}$) along it and stores $\zeta$ moles of solvent per unit length per unit thickness. The solvent flux $h_c$ thus characterizes flow that carries fluid from the crack to fill the newly forming fracture within the cohesive zone. The solvent flux must satisfy a mass conservation law in the form
\begin{align}
\label{eq:balasolcoh}
\dot{\zeta}=-\frac{\partial h_c}{\partial \hat{x}_1}-(h_2^+-h_2^-)\,,
\end{align}
where $h_2^+ = \mb{h}^+\cdot\mb{e}_2$ and $h_2^- = \mb{h}^-\cdot\mb{e}_2$ are the fluxes of solvent exchanged with the cohesive zone through the planes $\hat{x}_2 = 0^+$ and $\hat{x}_2 = 0^-$, respectively. Given the symmetry of the Mode I fracture problem studied here, $h_2^+=-h_2^-$.

Similarly to the swelling constraint for the gel, which prescribes a link between the volume change and the change in solvent concentration, we relate the opening $\delta=u_2^+-u_2^-=2u_2^+$ of the cohesive zone to the increment in solvent concentration $\zeta$ through the constraint $\delta = \Omega\zeta$. With this, we may eliminate $\zeta$ from the formulation of the problem and rewrite Eq.~\eqref{eq:balasolcoh} as 
\begin{align}
\label{eq:balasolcoh2}
\dot{\delta}+\Omega \frac{\partial h_c}{\partial \hat{x}_1}+2 \Omega h_2^+=0\,.
\end{align}

As concerns the constitutive laws, we begin by prescribing the following relation for the cohesive traction
\begin{align}
\label{eq:cohstress}
\sigma = \sigma^n(\delta) -\frac{\mu_c}{\Omega}\,, && 
\sigma^n(\delta) =
\begin{cases} 
\displaystyle \frac{\sigma_c}{\delta_f}\delta\,, & 0\leq \delta \leq \delta_f\,, \\
\displaystyle 0 \,, & \delta > \delta_f\,,
\end{cases}
\end{align}
where $\sigma_c$ is the cohesive strength of the polymer network, $\delta_f$ is the crack opening displacement at which the polymer network fails, whereas $\mu_c$ is the chemical potential of the solvent within the cohesive zone. We enforce chemical equilibrium between the solvent in the cohesive zone and that in the gel, \textit{i.e.} $\mu_c = \mu^+ = \mu^-$. Moreover, we relate $\sigma_c$ and $\delta_f$ to the fracture energy $\Gamma$ of the polymer network by requiring the work performed by $\sigma^n$ until failure to equal $\Gamma$, such that
\begin{align}
\label{eq:fracenergycoh}
\Gamma = \frac{1}{2}\sigma_c \delta_f .
\end{align}
The choice of a linear traction-separation law for $\sigma^n$ as in Eq.~\eqref{eq:cohstress} implies a theoretically infinite cohesive length. Nevertheless, we can give an operational definition of $R_v$ as the length across which crack opening varies from $\delta_f$ to $\delta_f/100$. As is clear from this definition, $R_v$ is determined by the solution of the problem.

Finally, for the tangential flux $h_c$ we enforce Darcy's law such that
\begin{align}
\label{eq:darcycoh}
h_c = -M_c\frac{\partial \mu_c}{\partial \hat{x}_1}\,,
\end{align}
with $M_c$ the solvent mobility within the cohesive zone. Given the absence of experimental evidence of how flow takes place relative to a polymer network that is tearing apart, we choose a linear model as the simplest possibility, such that $M_c=\delta_{f}M$.
 
We proceed now with our treatment by exploring the energetic aspects of fracture in the presence of the poroelastic cohesive zone. First, the flux of energy $\Phi$ in Eq.~\eqref{eq:powerbal} must be replaced by the power exchanged between the bulk gel and the cohesive zone, which may be evaluated as 
\begin{align}
	\label{eq:cohG}
	\Phi  = \int_{0}^{R_v}{(\sigma \dot{\delta}-2\mu_c h_2^+)\mbox{d}\hat{x}_1}.
\end{align}
Then, to relate this to the energy release rate $\mc{G}_v^\infty $ in the far field, we retrace the steps that led us to Eq.~\eqref{eq:Gexp} by making use of Eqs.~\eqref{eq:chempowdiv}-\eqref{eq:nonsteadypsi}, where the (vanishing) tip integrals must now be replaced by the following terms
\begin{align}
\label{eq:chempowdivcoh}
 -\int_{0}^{R_v}{2\mu_c h_2^+\,\mbox{d}\hat{x}_1}\,, && -\int_{0}^{R_v}{\sigma\delta'\,\mbox{d}\hat{x}_1}\,,
\end{align}
respectively. By doing so, after some calculations we obtain
\begin{align}
\label{eq:toughcoh}
\mc{G}_v^\infty = \Gamma-\int_{\delta_f}^{0}{\frac{\mu_c}{\Omega}\,\mbox{d}\delta}  + \int_{\mc{R}^\infty}{\frac{\mu}{\Omega} \frac{\partial \varepsilon}{\partial \hat{x}_1}}\,\mbox{d}A\,.
\end{align}
We observe that $\mu_c/\Omega$ will be negative in the cohesive zone due to the pressure gradient required to draw solvent into it. Therefore, we conclude that the second term on the right hand side of Eq.~\eqref{eq:toughcoh} contributes a positive increment to the toughness of the gel, in addition to the contribution provided by the third term on the right hand side. We note that Eq.~\eqref{eq:toughcoh} is valid for crack growth in general and is not confined solely to the case of steady state propagation, which will be the focus of the next section.  

\section{Steady-state crack propagation: numerical results and discussion}
\label{sec:results}

In this section, we investigate numerically the problem of steady-state crack propagation to quantify possible sources of toughening. As we have seen in Section~\ref{sec:ass}, this analysis will also provide leading-order results for transient crack propagation. In particular, we explore the effect of crack tip velocity and material parameters on the degree of toughening $K_v^\infty/K_\textrm{o}^\infty$, and compare the results obtained with and without the poroelastic cohesive zone model. To this aim, we carry out numerical simulations using the finite element method. Given the assumption of steady-state crack growth, we take a time-independent domain centered at the crack tip as the computational domain, with characteristic size $L\gg l$ for the small-scale process zone hypothesis to hold. This requirement implies, in general, large values of the local (mesh-dependent) Peclét numbers, and thus demands a numerical stabilization technique \citep{JohnsonBook} for the advection-diffusion equation \eqref{eq:balaeq}$_2$ with $\varepsilon'=0$, to avoid excessively fine meshes. Here, for the sake of simplicity, we adopt artificial diffusion stabilization by replacing the solvent flux $\mb{h}$ in the mass balance with $\bar{\mb{h}} = (1+\eta h v)h_1\mb{e}_1+h_2\mb{e}_2$, where $h$ is the local mesh size and $\eta$ is a small tuning parameter. For the symmetry of the problem at hand, we restrict the computations to the rectangular domain $\mc{B}$ in the half-plane $\hat{x}_2\geq 0$, with $-L/2\leq\hat{x}_1\leq L/2$ and $0\leq\hat{x}_2\leq L/2$.

As standard in finite element procedures, we first recast the balance equations \eqref{eq:linbal} in weak form:
\begin{align}
\label{eq:weakbalance}
&\int_{\mc{B}}{\mb{T}\cdot\nabla\tilde{\mb{u}}\,\mbox{d}A} = \int_{\partial\mc{B}}{\mb{t}\cdot\tilde{\mb{u}}\,\mbox{d}S}\,, && \int_{\mc{B}}{\mb{h}_t\cdot\nabla\tilde{\mu}\,\mbox{d}A} = 0\,,
\end{align}
where $\mb{t}=\mb{Tn}$ is the traction corresponding to the Mode I stress field Eq.~\eqref{eq:modeIsol} evaluated at the far-field boundary of $\mc{B}$, $\mb{h}_t = \Omega\bar{\mb{h}}-v\varepsilon\mb{e}_1$ is the total solvent flux, whereas $\tilde{\mb{u}}$ and $\tilde{\mu}$ are the virtual fields corresponding to the unknowns $\mb{u}$ and $\mu$ of the problem, respectively. Since the traction is zero on crack faces, $\partial \mc{B}$ includes only the far-field boundary of the computational domain. The system \eqref{eq:weakbalance} is solved together with the essential boundary conditions Eq.~\eqref{eq:pressinc} and $\mu=0$ on the crack face $(-L/2\leq\hat{x}_1\leq 0, \hat{x}_2 = 0)$, and the symmetry condition $u_2=0$ on $(0\leq \hat{x}_1\leq L/2, \hat{x}_2 = 0)$.
As usual, we assume that the virtual fields vanish on the boundaries where Dirichlet conditions are prescribed. The stress $\mb{T}$ and the flux $\mb{h}$ are expressed as a function of $\nabla\mb{u}$ and $\nabla\mu$, respectively, through the constitutive law of Eq.~\eqref{eq:const} and Darcy's law.

For the computations involving the poroelastic cohesive zone model, the virtual work of the cohesive tractions
\begin{align}
\label{eq:cohvirtwork}
&-\int_{0}^{L/2}{\sigma \tilde{u}_2\,\mbox{d}\hat{x}_1}\,,
\end{align}
with $\sigma$ given by Eq.~\eqref{eq:cohstress}, is added to the right hand side of Eq.~\eqref{eq:weakbalance}$_1$ and replaces the symmetry condition on $u_2$, while
\begin{align}
\label{eq:cohsolflux}
&-\Omega\int_{0}^{L/2}{h_2 \tilde{\mu}\,\mbox{d}\hat{x}_1}=-\frac{1}{2}\int_{0}^{L/2}{h_{ct}\frac{\partial \tilde{\mu}}{\partial \hat{x}_1}\,\mbox{d}\hat{x}_1}
\end{align}
contributes to the right hand side of Eq.~\eqref{eq:weakbalance}$_2$, with $h_{ct}=\Omega h_c - 2vu_2$. The latter equation is the steady-state, weak form of the balance of solvent mass~\eqref{eq:balasolcoh2}, where the solvent flux follows Eq.~\eqref{eq:darcycoh} with $\mu_c=\mu$. To avoid dealing with the unknown cohesive length $R_v$, we have extended the integrals in Eqs.~\eqref{eq:cohvirtwork}-\eqref{eq:cohsolflux} all along the semi-axis $\hat{x}_1\geq 0$. 
We have verified {\it a posteriori} that this modification does not significantly alter the solution in $R_v\leq\hat{x}_1\leq L/2$, where $\delta \approx h_c \approx h_2 \approx 0$.

\begin{figure}[!t]
	\centering
	\includegraphics[scale=1]{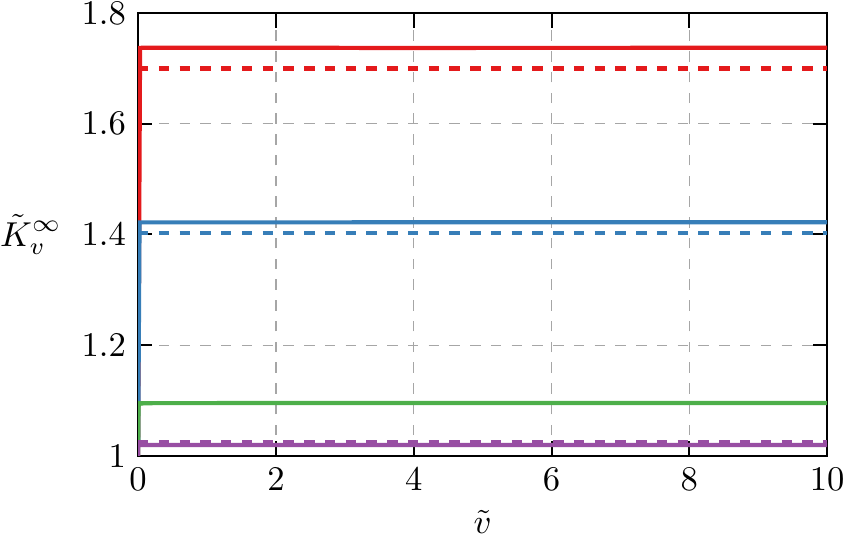}
	\caption{Toughness ratio $\tilde K_v^{\infty}$ as a function of the dimensionless crack tip velocity $\tilde v$. Numerical results (solid lines) for a steadily propagating crack and analytical estimates (dashed lines) from Eq.~\eqref{eq:Kinfty}, obtained using the energy release rate approach (no cohesive zone), with the fracture criterion $\mc{G}=\Gamma$. From top to bottom, lines correspond to increasing values of $\kappa/G = 1$ (red), 2 (blue), 10 (green), 50 (purple).}
	\label{fig:Rcurve}
\end{figure} 

In the \textit{energy release rate approach}, \textit{i.e.}~without the cohesive zone model, we enforce the crack propagation criterion $\mc{G} = \Gamma$ using an optimization procedure. Specifically, for each crack tip velocity $v$, this procedure solves for the applied stress intensity factor $K_v^{\infty}$ by minimizing the squared difference $(\mc{G}-\Gamma)^2$, where $\mc{G}$ is computed from Eq.~\eqref{eq:Jdomint}. We evaluate numerically the toughness ratio $\tilde K_v^\infty = K_v^\infty/K_\textrm{o}^\infty$, which, as dictated by dimensional analysis, is a function of two dimensionless groups $\kappa/G$ and $\tilde v = v \Gamma/(D_c G)$:
\begin{align}
	\tilde K_v^\infty = f\left(\frac{\kappa}{G},\tilde v\right)\,.
\end{align} 
The results from our computation are reported in Fig.~\ref{fig:Rcurve} and confirm the existence of a velocity independent toughening. These results closely agree with the estimate of $\tilde K_v^\infty$ given by Eq.~\eqref{eq:Kinfty}, where $\gamma \approx 3.4$ has been fitted to the numerical simulation for $\kappa/G=50$ and then kept constant as $\kappa/G$ varies.

We employ now the \textit{poroelastic cohesive zone model} as described in Section~\ref{sec:cohzone}. In this case, $K_v^\infty$ is found by a numerical optimization procedure that minimizes the squared difference $(\delta/\delta_f -1)^2$, where $\delta$ is evaluated at $\hat{x}_1=0$. Keeping the ratio of $M_c/(\delta_f M)$ fixed, we deduce that the toughness ratio depends now on the additional dimensionless group $\sigma_c/G$:
\begin{align}
	\tilde K_v^{\infty} = g\left(\frac{\kappa}{G},\tilde v,\frac{\sigma_c}{G}\right)\,.
\end{align}

\begin{figure}[!h]
	\centering
	\includegraphics[scale=1.0]{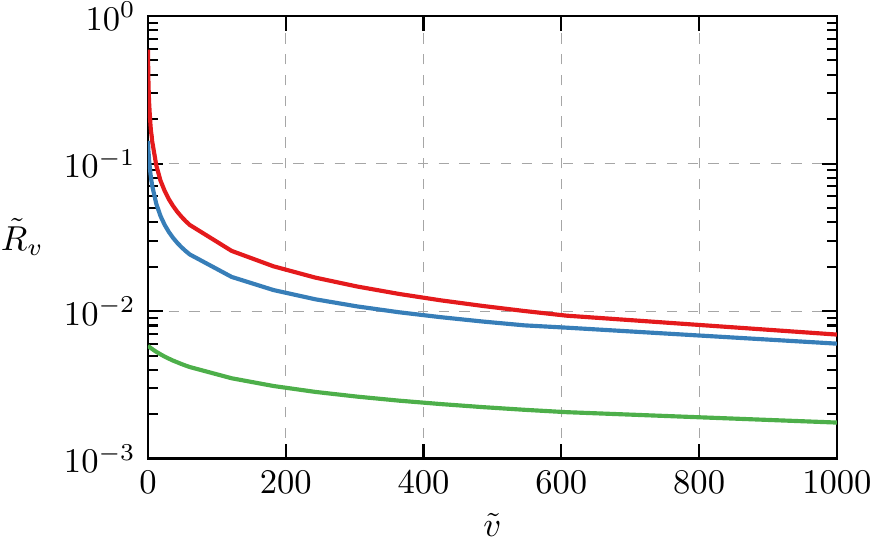}
	\caption{Dimensionless cohesive length $\tilde{R}_v = R_v  G/\Gamma$ as a function of the dimensionless crack tip velocity $\tilde{v}$ for $\kappa/G = 10$. From top to bottom, lines correspond to increasing values of $\sigma_c/G = 100$ (red), 200 (blue), 1000 (green).}
	\label{fig:cohlvsv}
\end{figure}

For the accuracy of the numerical simulations, it is important to ensure that all the relevant fields are appropriately resolved. We have verified numerically that the cohesive length $R_v$ significantly decreases with both crack tip velocity and the ratio $\sigma_c/G$, see Fig.~\ref{fig:cohlvsv}. Thus, we have checked that in all computations at least 10 finite elements span the cohesive length. In the same spirit, we have also chosen the size $L$ of the computational domain to be large with respect to $R_v$.

\begin{figure}[!h]
	\centering
	\includegraphics[scale=1]{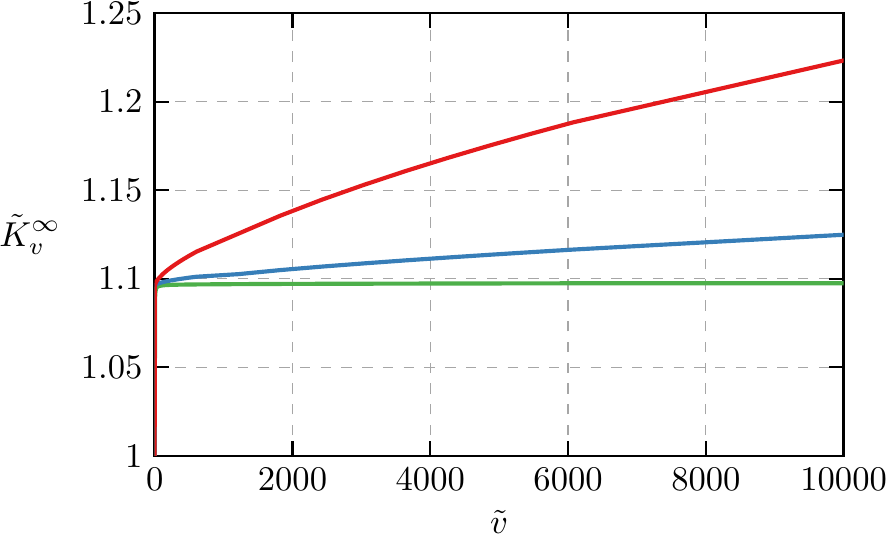}
	\caption{Toughness ratio $\tilde K_v^{\infty}$ as a function of the dimensionless crack tip velocity $\tilde v$. Numerical results for a steadily propagating crack, obtained using the cohesive zone model described in Section~\ref{sec:cohzone} with $\kappa/G=10$. From top to bottom, lines correspond to increasing value of $\sigma_c/G = 100$ (red), 200 (blue), 1000 (green).}
	\label{fig:Rcurvecoh}
\end{figure}

\begin{figure}[!h]
	\centering
	\includegraphics[scale=1]{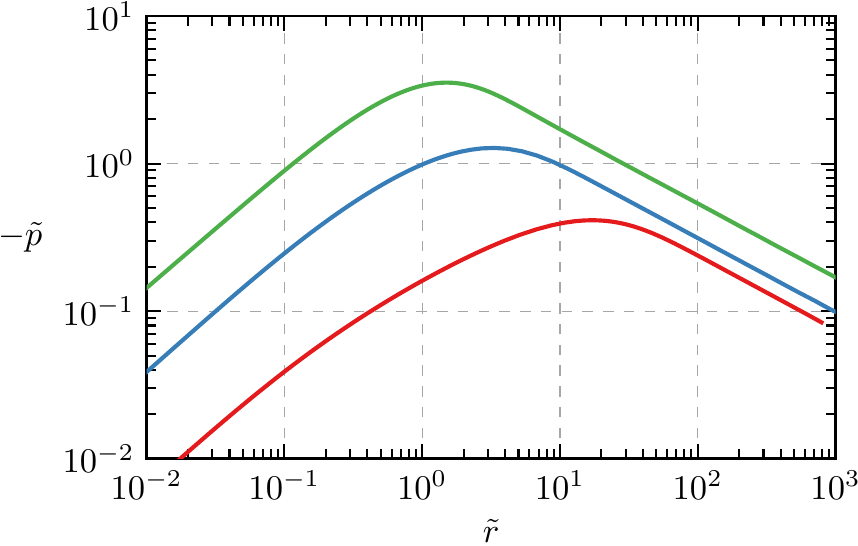}
	\caption{Dimensionless negative solvent pressure $-\tilde{p} = -\mu/(\Omega G)$ ahead of the crack tip as a function of the dimensionless coordinate $\tilde{r}=r/R_v$, with $\kappa/G = 10$ and $\sigma_c/G = 200$. From bottom to top, lines correspond to increasing values of the dimensionless velocity $\tilde v = 0.6$ (red), 6 (blue), 60 (green).}
	\label{fig:adimpresscoh}	
\end{figure}

The results for the toughness ratio based on the cohesive zone model are shown in Fig.~\ref{fig:Rcurvecoh} for a fixed value of $\kappa/G$ and for several values of $\sigma_c/G$. In sharp contrast to the fracture criterion based on crack tip toughness, recall Fig.~\ref{fig:Rcurve}, now the effective toughness of the gel increases with crack tip velocity. We have already anticipated that the solvent pressure within the cohesive zone provides a toughening contribution that was not captured by the energy release rate approach, recall Eq.~\eqref{eq:toughcoh}. The dependence of this contribution on crack tip velocity may be explained by considering the concurrent evolution of solvent pressure profile and cohesive length. As depicted in Fig.~\ref{fig:adimpresscoh}, an increase in crack tip velocity results in increased solvent pressure magnitude within the cohesive zone. As a consequence, the second term in Eq.~\eqref{eq:toughcoh} increases with $\tilde v$ and thus determines the observed trend of $\tilde K_v^\infty$.

Proceeding to consider the dependence of toughening on the ratio $\sigma_c/G$, we observe that, as the size of the cohesive zone vanishes, the solution obtained with the cohesive zone model recovers that obtained with the energy release rate approach (compare the corresponding lines in Fig.~\ref{fig:cohlvsv} and in Fig.~\ref{fig:Rcurvecoh}). Finally, for a given value of $\tilde v$, the toughening increases with cohesive length $R_v$ by virtue of the increased extent of the region where work must be done on solvent pressure.

\section{Conclusions and outlook}

In this study, the Mode I fracture toughness has been explored for a polymer gel specimen containing a semi-infinite, propagating crack. First, by adopting a fracture criterion that relies on crack tip toughness only, we have demonstrated the existence of a crack tip velocity-independent enhancement in the resistance to fracture of the gel. Physically, this toughening arises from the work performed by the solvent pressure against volume expansion within the process zone. The increase in effective toughness of the system, as quantified by the increase of the remotely applied stress intensity factor, has been evaluated through both asymptotic analyses and refined numerical simulations. 

Then, we have introduced a poroelastic cohesive zone model, which opens the way to the future study of the micromechanics of diverse fracture processes in polymer gels and to future numerical simulations of transient crack propagation. In contrast to the first approach, we have shown that, in the presence of a cohesive zone, the applied stress intensity factor needed to sustain crack growth increases with crack tip velocity. This behaviour is determined by an additional contribution to the fracture toughness that arises from the resistance to crack opening offered by the negative solvent pressure within the cohesive zone.

Our results highlight the importance of poroelasticity and, more generally, of the accurate description of the processes occurring in the fracturing region in quantifying possible toughening mechanisms in polymer gels. Besides their relevance in fracture of these synthetic materials, our findings may have a biological reach because many soft biological materials, for example brain tissue \citep{franceschini_2006}, have been shown to follow poroelasticity. Experiments should be carried out to confirm the outcomes of the present study and to provide further insight to improve the cohesive zone modeling here presented. Future work will also include the extension of our results to finite deformation poroelasticity.

\section*{Acknowledgments}
GN, AL and ADS acknowledge support from the European Research Council through AdG-340685--MicroMotility. Work by RMM  was supported by the U.S. Department of Energy (DOE), Office of Science, Basic Energy Sciences (BES) under Award \# DE-SC0014427.



\section*{References}

\end{document}